# Pilot Contamination Elimination for Channel Estimation with Complete Knowledge of Large-Scale Fading in Downlink Massive MIMO Systems

Qazwan abdullah[1], Norsaliza Abdullah[1], Adeb Salh[1], Lukman Audah[1], Nabil Farah[2], Abbas Uğurenver[3], Abdu Saif[4]

[1]Faculty of Electrical and Electronic Engineering, Universiti Tun Hussein Onn Malaysia
Johor, Malaysia
[2]Faculty of Electrical Engineering, Universiti Teknikal Malaysia Melaka, Melaka, Malaysia
[3]Faculty of Electrical Engineering İstanbul Aydın University, Istanbul, Turkey
[4]Faculty of Electrical Engineering Department, University of Malaya, Kuala Lumpur, Malaysia

[1]gazwan20062015@gmail.com

*Abstract*—Massive multiple–input multiple-output (MIMO) is a very important technology for future fifth generation systems (5G). However, massive MIMO systems are still limited because of pilot contamination (PC), which impacts the data rate (DR) due to the non-orthogonality of pilot sequences transmitted by users in the same cell to the neighbouring cells. We propose a channel estimation with complete knowledge of large-scale fading by using an orthogonal pilot reuse sequence (PRS) to eliminate PC in edge users with poor channel quality based on the estimation of large-scale fading and performance analysis of maximum ratio transmission (MRT) and zero forcing (ZF) precoding methods. We derived the lower bounds on the achievable downlink DR and signal-to interference noise ratio (SINR) based on assigning PRS to a user grouping that mitigated this problem when the number of antenna elements $M \to \infty$. The simulation results showed that a high DR can be achieved due to better channel estimation and reduced performance loss.

*Index Terms*—Massive MIMO, fifth generation (5G), signal-to-interference-noise ratio, maximum ratio transmission.

## I. INTRODUCTION

MASSIVE MIMO is a very important technology for interference reduction between neighbouring cells and multi-user interference to enhance the network capacity of 5G. In multi-cell massive MIMO systems, PC is a fundamental problem, which impacts the DR. Nevertheless, channel estimation can be used in time division duplex (TDD) for training the channel by transmitting PRS, where the PRS between neighboring cells increases the gain and avoids interference between adjacent cells [1-5]. PC occurs when there is a transmission of the same (orthogonal) pilot sequences from all users (UEs) at the same time and from the same cell to all neighbouring cells. Many authors have used different methods to eliminate PC. The author in [6-11] studied the effect of shifting pilots between neighbouring cells by transmitting signals at the same time, which provided better channel estimation and improved the operation of precoding. The other author [12-17] studied the reduction of PC based on orthogonal pilots between neighbouring cells and obtained the optimal pilot sequences for the set of all users. While, the author in [4-22] studied the effects of PC in both uplink and downlink and with and without pilot aided in a practical massive MIMO using linear precoding in the maximum ratio combination and ZF to achieve high DR. In this paper, we propose channel estimation with complete knowledge of large-scale fading by using PRS to eliminate PC in edge users. Therefore, we estimated the large-scale fading and obtained better performance analysis based on the MRT and ZF precoding when the number of antenna elements $M$ and number of users (UEs) $K$ increased to large numbers.

*Notations:* We use $\mathcal{CN}(.,.)$ for the circularly symmetric complex Gaussian distribution. $[\cdot]$ and $var(.)$ stand for the expectation and variance operations, respectively. The Hermitian transpose is denoted by $(.)^M$.

## II. SYSTEM MODEL

In downlink multi-cell massive MIMO systems, the base station (BS) transmits signals to UEs, in which here every cell has the $M$ antenna of BS to serve $K$, $(M \gg K)$ and operates at the same frequency. We assumed that the channel reciprocities were the same in the uplink and downlink (DL), where the channel model is independent and identically distributed (i.i.d.) for the correlated Rayleigh fading channel matrix and the properties minimum mean square error (MMSE) for channel estimation $h_{ljk} \in \mathbb{C}^{M \times 1}$ could assign a different antenna correlation to every channel between the users in the BS $l$ and $M$ in the BS $j$. In $\theta_{ljk} \in \mathbb{C}^M$, $M \times 1$ is the small-scale fading channel and $\psi_{ljk} \in \mathbb{C}^{M \times M}$ accounts for the corresponding channel correlation matrix for large-scale fading. In $[D_{lj}]_{k,k} =$

$\sqrt{\psi_{ljk}}$, $D_{ljk}$ is the diagonal matrix whose diagonal elements are $\sqrt{\psi_{lj}} = [\sqrt{\psi_{lj1}},..,\sqrt{\psi_{ljK}}]$. The channel between BS $l$ and the $Kth$ user in cell $j$ is given by

$$h_{ljk} = \sqrt{\psi_{ljk}}\,\Theta_{ljk} \quad (1)$$

We assumed that the BS was working with imperfect channel state information (CSI). The received signal $z_{jk}$ of the $Kth$ user inside the $jth$ cell can be expressed as

$$z_{jk} = \underbrace{\sqrt{\rho_d}h_{jjk}^H x_{jk}a_{jk}}_{desired\ signal} + \underbrace{\sqrt{\rho_d}\sum_{i=1,i\neq k}^{K} h_{jjk}^H x_{ji}a_{ji}}_{intra-cell\ interference} + \underbrace{\sqrt{\rho_d}\sum_{l=1,l\neq j}^{L}\sum_{i=1}^{K} h_{ljk}^H x_{lk}a_{lk}}_{inter-cell\ interference} + n_{jk} \quad (2)$$

where $h_{ljk}^H$ is the Hermitian transpose channel matrix and the UEs $K$ in every cell use orthogonal pilot are reused to estimate the channel. $h_{jjk} = [h_{jj1}\ldots\ldots h_{jjK}] \in \mathbb{C}^{M \times K}$ represents the downlink channel between BS $j$ and user $K$ in its cell. The transmit signal vector of BS is $y_{lk} = x_{lk}a_{lk} \in \mathbb{C}^M$, $a_{lk} \in \mathbb{C}^{M \times K}$ is the linear precoding matrix, $x_{lk} \in \mathbb{C}^K \sim \mathcal{CN}(0, I_K)$ is the data symbol transmitted from BS in cell $l$ to the UEs, $\rho_d$ is the DL transmit power, and $n_{jk} \sim \mathcal{CN}(0_{M\times 1}, I_M)$ is the received noise vector.

## A. Channel Estimation with complete knowledge of large-scale fading

We used channel estimation in the downlink by exploiting channel reciprocity in the TDD. Consequently, due to the reciprocity of electromagnetic waves at the transmit signal in the uplink, we assumed that channel reciprocity could be estimated by corresponding the UEs in cell $j$ in downlink channel $H_{jjk}^H$, which is the Hermitian transpose of uplink $H_{jjk}$. The received training signal in the reuse same pilot sequence for neighboring cells could be used to estimate the channel as

$$£_{jk} = h_{jjk} + \sum_{l\neq j}^{K} h_{ljk} + \frac{n_{jk}}{\rho^{1/2}} \quad (3)$$

where $\rho$ is the length of the training transmit power, which is proportional with the effective pilot SNR. By using the correlated received pilot matrix, we obtained the conventional (low complexity) channel estimation, where $\hat{h}_{jjk}$ is convenient for a large number of $M$ [5],[23],[26] at multiple $\hat{h}_{jjk}$ by $£_{jk}^{tr}$. We assumed that the low complexity of channel estimation for $\hat{h}_{jjk}$ could be written as

$$\hat{h}_{jjk} = \psi_{jjk}\varphi_{jk}\sum_{l\neq j}^{K} h_{ljk} + \frac{n_{jk}}{\rho^{1/2}} \quad (4)$$

## B. Downlink Transmission

Suppressing PC depended on increasing the number of transmit pilot reuses with large-scale fading. Furthermore, the orthogonal pilot sequences in the multi-cell needed symbols $K \times L$. Therefore, we used long training sequences, where the variance channel is written as

$$\varphi_{jk} = \left(\frac{I_M}{\rho} + \sum_{l=1}^{L}\psi_{ljk}\right)^{-1} \quad (5)$$

The increase of the orthogonal pilot sequence and the number of UEs inside the cell led to a pilot sequence of $Y_p = K = \infty$. During the uplink training phase, $K$ users in the same cell transmitted the orthogonal pilot sequence, where $Y_p$ is the symbol of the pilot sequence in the downlink that appeared for the desired channel; the received signal at the BS is given by

$$\xi_{ljk} = \sqrt{\rho_d Y_p}\sum_{l=1}^{L} h_{ljk} + n_{lk} \quad (6)$$

where $Y_p$ is the pilot sequence in each coherence interval that could be allocated between the uplink and downlink of the transmit data signal. From the properties of MMSE for the estimation channel $h_{\widetilde{ljk}} = \hat{h}_{jlk} - h_{ljk}$ the channel estimation between users $K$ and BS cells $L$ was found, where $h_{jjk} \sim \mathcal{CN}(0, \varphi_{jk})$ is the uncorrelated estimation independent of $h_{\widetilde{jjk}} \sim \mathcal{CN}(0, (\varphi_{jk} - O_{ljk})I_M)$, and $O_{ljk} = \psi_{jjk}\varphi_{jk}\psi_{ljk}$, which is also independent of antenna $M$ for large-scale fading [6-10],[22-28]. The MMSE of the Kth UEs can be expressed as

$$\tilde{h}_{ljk} = \mathbb{E}\|\hat{h}_{ljk} - h_{ljk}\|^2$$
$$= \mathbb{E}\{|\varphi_{jk}(\xi_{ljk} +) - h_{ljk}|^2\}$$
$$= \mathbb{E}\{|(\varphi_{jk} - I_M)(\xi_{ljk}) + n_{lk}|^2\}$$
$$= \{\psi_{ljk}(\varphi_{jk} - I_M)(\varphi_{jk} - I_M)^H(\xi_{ljk}) + \sigma^2 I_M\}$$
$$\tilde{h}_{ljk} = \{\psi_{ljk}(\varphi_{jk} - I_M)(\varphi_{jk} - I_M)^H(\Theta_{ljk}(\rho_d Y_p \sum_{l=1}^{L}\psi_{ljk})\Theta_{ljk}^H)\xi_{ljk} + \sigma^2 I_M\}$$
$$\tilde{h}_{ljk} = \{\psi_{ljk}((\frac{I_M}{\rho}+\sum_{l=1}^{L}\psi_{ljk})^{-1} - I_M)((\frac{I_M}{\rho}+\sum_{l=1}^{L}\psi_{ljk})^{-1} - I_M)^H((\Theta_{ljk}(\rho_d Y_p \sum_{l=1}^{L}\psi_{ljk})\Theta_{ljk}^H)\xi_{ljk}) + \sigma^2 I_M\} \quad (7)$$

The MMSE estimate of a channel depends on $\xi_{ljk}$, which can be simplified by using matrix inversion where $\Theta_{ljk}\Theta^H_{ljk} = 1$ and the same vector $\Theta^H_{ljk}\xi_{ljk}$ for every channel estimation is proportional to the properties of MMSE for a BS. According to the training received signal for large-scale fading $\Theta^H\xi_{ljk}$, the estimation channel proportional with MMSE is

$$\tilde{h}^H_{ljk}/\|\tilde{h}_{ljk}\| = \frac{\Theta^H_{ljk}\xi_{ljk}}{\|\Theta^H_{ljk}\xi_{ljk}\|} \quad (8)$$

Using equation (8), the effect of PC due to the correlated precoding channel matrix and UEs $K$ in different cells during signal transmission from BS to the UEs is $\mathbb{E}[h_{ljk}a_{lk}] = \mathbb{E}\|\tilde{h}^H_{ljk}\|$. Channel estimation $\tilde{h}_{ljk}$ is related to the channel response $h_{ljk}$ based on equations (4) and (8), which can be expressed as

$$\tilde{h}_{ljk} = \frac{\rho_d Y_p \psi_{ljk}}{1+\rho_d Y_p \sum_{l=1}^{L}\psi_{lik}}\Theta^H_{ljk}\xi_{ljk} \quad (9)$$

## C. Achievable Data Rate

In this section, we present our analysis of the achievable DR based on effective SINR, which is an important method to determine system performance. The problem of linear precoding is due to the received training signal and different numbers of antennas at the BS, which cannot be formulated directly [29-33]. The MRT precoding maximized the SNR of every UE $K$ for the desired signal under constrained power, in addition to the MRT, largely from the interference of $\mathcal{A}_j =$

$[a_{1,MRT},..,a_{K,MRT}] = [\tilde{h}^H_{lj1},..,\tilde{h}^H_{ljk}] = \tilde{H}^H_{ljk}$. ZF precoding could schedule up to $K$ users, send one data stream per user, and enhance processing to reflect the impact of background noise and unidentified user interference. In addition, the DR was based on the condition of channel inverse $a_j = \tilde{H}_{ljk}(\tilde{H}^H_{ljk}\tilde{H}_{ljk})^{-1} / \|\tilde{H}_{ljk}(\tilde{H}^H_{ljk}\tilde{H}_{ljk})^{-1}\|^2$. We derived the lower bound of the achievable DR as:

$$\mathcal{R}_{jk} = \sum_{l=1}^{L}\sum_{i=1}^{K}(1-\partial)\left[\log_2(1+\Gamma^{dl}_{jk})\right] \quad (10)$$

where $(1-\partial)$ represents the loss of pilot signaling for the pre-log factor and $0 < \partial < 1$ is used to evaluate the achievable DR. The channel response was estimated based on mitigated PC using a correlated channel matrix. We decomposed the received signal as

$$\mathcal{Y}_{jk} = \sqrt{\rho_d \Upsilon_p}\mathbb{E}\{h^H_{jjk}a_{jk}\}x_{jk} + \sqrt{\rho_d \Upsilon_p}\sum_{i=1,i\neq k}^{K}(h^H_{jjk}a_{jk} - \mathbb{E}\{h^H_{jjk}a_{jk}\})\mathcal{V}_{jk} + \sqrt{\rho_d \Upsilon_p}\sum_{l=1,l\neq j}^{L}\sum_{i=1}^{K}h^H_{ljk}a_{lk}x_{lk} + n_{jk} \quad (11)$$

The achievable DR of the transmission from BS $l$ was obtained by considering Gaussian noise as the worst case of the uncorrelated noise channel, which can be written as

$$R_t = \sum_{l=1}^{L}\sum_{i=1}^{K}(1-\partial)\log_2\left(1+\frac{\mathbb{E}|Desired\ signal|^2}{\mathbb{E}|Uncorrelated\ noise|^2}\right) \quad (12)$$

The PRS was correlated between the channel $h^H_{ljk}$ and precoding $a_{lk}$ in a neighboring cell. We derived the SINR at the $kth$ UEs[34-37]; the desired signal for SINR can be expressed as

$$\mathbb{E}|Desired\ signal|^2 = \rho_d \Upsilon_p |\mathbb{E}[h^H_{jjk}a_{jk}]|^2 \quad (13)$$

From the uncorrelated noise, power can be expressed as

$$\mathbb{E}|Uncorrelated\ noise|^2 = \rho_d \Upsilon_p \mathbb{E}\{|h^H_{jjk}a_{jk} - \mathbb{E}[h^H_{jjk}a_{jk}]|^2\} + \sum_{i=l}^{L}\rho_d \Upsilon_p \mathbb{E}\{|h^H_{jjk}a_{jk}|^2\} + \rho_d \Upsilon_p \sum_{l=1,l\neq j}^{L}\sum_{i=1}^{K}\mathbb{E}\{|h^H_{ljk}a_{lk} - \mathbb{E}[h^H_{ljk}a_{lk}]|^2\} + \sigma^2 \quad (14)$$

$$\mathbb{E}|Uncorrelated\ noise|^2 = \rho_d \Upsilon_p \sum_{l=1,l\neq j}^{L}\sum_{i=1}^{K}\mathbb{E}[|h^H_{ljk}a_{lk}|^2] - \sum_{i=l}^{L}\rho_d \Upsilon_p |\mathbb{E}[h^H_{jjk}a_{jk}]|^2 + \sigma^2 \quad (15)$$

To get outstanding results of SINR with large numbers of $M$, we derived the SINR at the $kth$ UEs under MRT and ZF precoding, which can be expressed as

$$\Gamma^{dl}_{jk} = \frac{\rho_d \Upsilon_p |\mathbb{E}[h^H_{jjk}a_{jk}]|^2}{\rho_d \Upsilon_p \sum_{l=1,l\neq j}^{L}\sum_{i=1}^{K}\mathbb{E}[|h^H_{ljk}a_{lk}|^2] - \sum_{i=l}^{L}\rho_d \Upsilon_p |\mathbb{E}[h^H_{jjk}a_{jk}]|^2 + \sigma^2} \quad (16)$$

From the received signal, we estimated the average channel and interference power in the closed form for MRT precoding, which can be expressed as

$$\rho_d|\mathbb{E}[h^H_{ljk}a_{lk}]|^2 = \frac{d}{M\,var(h_{ljk})}\left|\left\{\mathbb{E}\|\tilde{h}_{ljk}\|^2\right\}\right|^2 \quad (17)$$

The noise variance was created from the large fading. We used the properties of the covariance channels matric $|\mathbb{E}[h^H_{ljk}a_{lk}]|^2$ from Kay [6]: $\rho_d|\mathbb{E}[h^H_{ljk}a_{lk}]|^2 = Mvar(h_{ljk})$. Using equation (4), we estimated the row vector of channel response $h_{ljk}$ between users in the $jth$ cell and the BS in the $lth$ cell, where $var(h_{ljk}) = (\rho_d\Upsilon_p\psi_{jjk}/(1+\rho_d\Upsilon_p\sum_{i=1}^{L}\psi_{lik}))$. The channel CSI was performed for MMSE when the diagonals of the covariance matrices were linearly independent [36-40]. From (14) the effect of PC for interference can be reduced by using correlated channel in the same cell to neighboring cells in (9). The SINR for MRT can be written as

$$\Gamma^{dl-mrt}_{jk} = \frac{Mvar(\tilde{h}_{jjk})}{Mvar(\tilde{h}_{ljk})+\sum_{l=1,l\neq j}^{L}\sum_{i=1}^{K}var(\tilde{h}_{jjk})+\frac{\sigma^2}{\rho_d}} \quad (18)$$

Based on ZF precoding $a_{lk} = \mathbb{E}|\tilde{H}_{ljk}(\tilde{H}^H_{ljk}\tilde{H}_{ljk})^{-1}| / \mathbb{E}\{\|\tilde{H}_{ljk}(\tilde{H}^H_{ljk}\tilde{H}_{ljk})^{-1}\|^2\}^{1/2}$, the inter-user interference could be suppressed because the user could be estimated directly without interference from other users with orthogonal PRS $a_{lk} = (M-K)var(\tilde{h}_{ljk})^{1/2}\tilde{H}_{ljk}(\tilde{H}^H_{ljk}\tilde{H}_{ljk})^{-1}$. Depending on a number of $M \to \infty$, the variance channel is $var\{\tilde{h}^H_{ljk}a_{jk}\}\xrightarrow[M\to\infty]{} 0$. From the correlation of channel property, the ZF precoding could be evaluated based on channel desired and channel estimation, and it could be computed as

$$\mathbb{E}\left\{\|\tilde{H}_{ljk}(\tilde{H}^H_{ljk}\tilde{H}_{ljk})^{-1}\|^2\right\}^{1/2} = \mathbb{E}\left[(\tilde{H}^H_{ljk}\tilde{H}_{ljk})^{-1}\right] = \frac{1}{(M-K)\,var(\tilde{h}_{ljk})} \quad (19)$$

According to the numerator and denominator of (16), we can simplify the linear precoding of ZF as

$$\rho_d|\mathbb{E}[h^H_{ljk}a_{lk}]|^2 = \rho_{jk}(M-K)\,var(\tilde{h}_{ljk}) \quad (20)$$

The interference in the denominator can be written as

$$\rho_d\sum_{l=1,l\neq j}^{L}\sum_{i=1}^{K}\mathbb{E}[|h^H_{ljk}a_{lk}|^2] - \rho_{jk}|\mathbb{E}[h^H_{jk}a_{jk}]|^2 + \sigma^2 = \rho_d(M-K)\,var(\tilde{h}_{ljk}) + \sum_{l=1,l\neq j}^{L}\sum_{i=1}^{K}\rho_d var(\tilde{h}_{ljk}) + \sigma^2 \quad (21)$$

The lower bound for large scale fading channel, by substituting (20) and (21) into (16), when the BS used ZF precoding, the SINR is

$$\Gamma^{dl-zf}_{jk} = \frac{(M-K)var(\tilde{h}_{jjk})}{(M-K)var(\tilde{h}_{ljk})+\sum_{l=1,l=j}^{L}0_{jk}+\sum_{l=1,l\neq j}^{L}\sum_{i=1}^{K}var(\tilde{h}_{jjk})+\frac{\sigma^2}{\rho_d}} \quad (22)$$

### D. Our Proposal

In equation (10), the achievable SINR is proportional to large-scale fading $\psi_{ljk}$ according to the location of $K$ in every cell. Therefore, sorting the UEs in a cell according to the best and worst channel conditions requires accurate channel quality of the large-scale fading $\psi_{ljk}$ of UEs in $ith$ cells when $j \neq l$. We improved the channel quality of users by assigning orthogonal PRS to the edge user group to mitigate pilot contamination in equations (18) and (22) based on large-scale fading $\psi_{jlk}$ and dividing users into two group as

$$\psi_{ljk} = \ell_i \geq \tau\mu_i \to \begin{cases} Yes \to & center\ users \\ No \to & edge\ users \end{cases} \quad (23)$$

where the grouping parameter $\tau$ can be chosen in the edge users to improve the DR, $\tau\mu_i$ represents grouping threshold value, and $\ell_i$ is the user's channel quality corresponding to $\psi_{ljk}$. Therefore, good channel quality can be obtained by reusing the same PRS to center users while assigning orthogonal PRS to the edge users [12], [19], [28]. The user grouping $\mu_i$ can be written as

$$\mu_i = \sum_{k=1}^{K} \frac{\max[\mathcal{b}_{i1}, \mathcal{b}_{i2},\dots,\mathcal{b}_{iK}]+\min[\mathcal{b}_{i1}, \mathcal{b}_{i2},\dots,\mathcal{b}_{iK}]}{2} \quad (24)$$

where $K_{ic} = \text{card}[k:\dots, \mathcal{b}_i > \mu_i]$ represents the number of center users and $K_{ie} = \text{card}[k:\dots, \mathcal{b}_i \leq \mu_i]$ represents the number of edge users. From the hexagonal cells, the $ith$ cell supports $K_i$ users and the orthogonal of PRS is based on user grouping

$$K_i = \frac{K_{ic}+LK_{ie}}{K} = \frac{\partial}{K}(\sum_{l=1}^{L} LK_{ie} + \max[K_{ic}, K_{ic}\dots, K_{ic}]) \quad (25)$$

The achievable high DR based on mitigated PC was due to the correlated precoding vector matrix in the same cell and the channel user in neighboring cells. The achievable DR with ZF and MRT precoding is given as

$$\mathcal{R}^{mrt,\ zf}{}_{jk} = \sum_{l=1}^{L}\sum_{i=1}^{K}\left(1 - \frac{\partial}{K}(\sum_{l=1}^{L} LK_{ie} + \max[K_{ic}, K_{ic}\dots, K_{ic}])\right)\log_2(1 + \Gamma_{jk}^{dl-mrt,zf}) \quad (26)$$

In equation (26), some losses occurred in the DR due to large-scale fading, which can be avoided by increasing the number of antenna $M$ and ensuring a uniform movement of location users $K$.

III. NUMERICAL RESULTS

In this section, we present the numerical results using Monte-Carlo simulations.

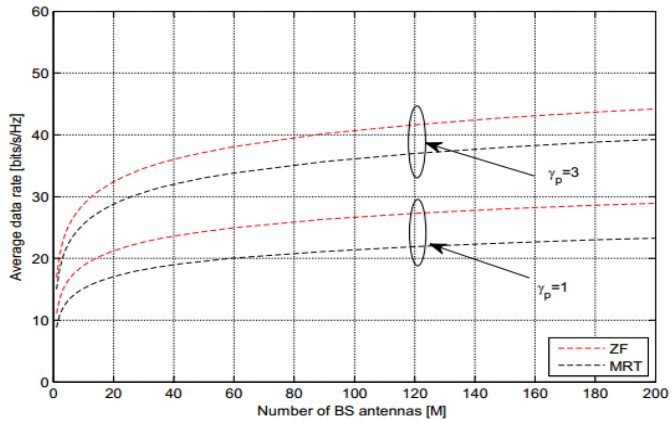

Fig. 1. Achievable data rate with number of antennas $M$.

Fig. 1 shows that at higher spatial resolutions, a uniform number of users inside every cell and channel estimation provided complete knowledge of large-scale fading. These made the target cell have achievable high DR when the grouping parameter $\tau$ increased with $M = 256$ and $K = 10$. Fig. 1 also shows that the increase of PRS from 1 to 3 increased the achievable DR. Furthermore, the ZF precoding provided more DR than MRT for large pilot reuse, because the ZF was able to work at high SINR. Moreover, the achievable DR can be improved if the number of edge users or grouping parameter $\tau$ is selected. Finally, the increased numbers of antennas in the multi-cell made these channels be orthogonal to others and suppressed the interference between neighboring cells.

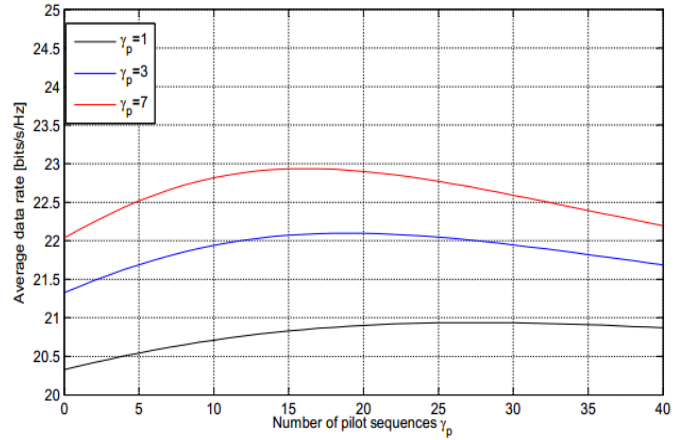

Fig. 2. Average data rate with number of pilot sequences $\Upsilon_p$.

According to equation (26), high achievable DR can be obtained, depending on user grouping, to mitigate PC in edge users if the number of edge users is selected. Moreover, in the case of the number of $\Upsilon_p = 7$, the system performance resulted in high performance of DR, compared to $\Upsilon_p = 1$. Fig. 2 shows that the number of PRS did not increase arbitrarily in order to decrease the interference due to the PC employed form channel estimation in all BS, which caused the critical limitation in coherence channel estimation and decreased the capacity of the DR. Fig. 2 shows that the average DR started to increase and then gradually decreased, which means that the increased DR of edge users becomes larger than the decreased DR of center users.

IV. CONCLUSIONS

This paper presented the analysis of MRT and ZF precoders with PC in a downlink multi-cell massive MIMO system. Due to channel estimation, this required comparing the received pilot from every UE with the recognized pilot signal associated with that UE. Channel estimation accuracy evaluated at the received signal and suppressed PC depended on orthogonal PRS for the edge users in the neighboring cells with large-scale fading. The numerical results showed that the use of orthogonal pilot reuse in the downlink reduced performance loss, provided better estimation quality in the channel, and maximized the high data rate.